\documentclass[letterpaper]{article} 
\usepackage{natbib} 
\usepackage{aaai2026} 
\usepackage{times} 
\usepackage{helvet} 
\usepackage{courier}  
\usepackage[hyphens]{url}  
\usepackage{graphicx}
\usepackage{caption} 
\usepackage{algorithm}
\usepackage{algorithmic}
\usepackage{amsmath}
\usepackage{booktabs} 
\usepackage{multirow}
\usepackage{enumitem}
\usepackage{amssymb} 
\usepackage{siunitx}
\usepackage{makecell}
\usepackage{marvosym}

\usepackage{makecell}

\usepackage[table]{xcolor}

\usepackage{bibentry}

\usepackage[most]{tcolorbox}
\usepackage{newfloat}
\usepackage{listings}
\DeclareCaptionStyle{ruled}{labelfont=normalfont,labelsep=colon,strut=off} 
\floatstyle{ruled}
\newfloat{listing}{tb}{lst}{}
\floatname{listing}{Listing}
\title{ChartEditor: A Reinforcement Learning Framework for Robust Chart Editing}

\author{
    Liangyu Chen\textsuperscript{\rm 1}\equalcontrib, Yichen Xu\textsuperscript{\rm 1}\equalcontrib, Jianzhe Ma\textsuperscript{\rm 1}, Yuqi Liu\textsuperscript{\rm 2}, Donglu Yang\textsuperscript{\rm 1}, Liang Zhang\thanks{Independent Researcher.}\\
    Zihao Yue\textsuperscript{\rm 1}, Wenxuan Wang\textsuperscript{\rm 1}\thanks{Qin Jin and Wenxuan Wang are the corresponding authors.}, Qin Jin\textsuperscript{\rm 1}\footnotemark[3]
}
\affiliations{
    \textsuperscript{\rm 1}Renmin University of China,     \textsuperscript{\rm 2}Chinese University of Hong Kong\\
    wenxuanwang@ruc.edu.cn, qjin@ruc.edu.cn

}

\newtcolorbox{blackpromptbox}[1][]{%
breakable,
colback=gray!10,
colframe=black,
fontupper=\small,
left=0.5mm, right=0.5mm, top=1mm, bottom=1mm,
boxrule=0.8pt,
sharp corners,
title={Prompt},
fonttitle=\bfseries,
#1
}
\begin{document}

\maketitle

\begin{abstract}
Chart editing reduces manual effort in visualization design. Typical benchmarks limited in data diversity and assume access to complete chart code, which is seldom in real-world scenarios. To address this gap, we present ChartEditVista, a comprehensive benchmark consisting of 7,964 samples spanning 31 chart categories. It encompasses diverse editing instructions and covers nearly all editable chart elements. The inputs in ChartEditVista include only the original chart image and natural language editing instructions, without the original chart codes. ChartEditVista is generated through a fully automated pipeline that produces, edits, and verifies charts, ensuring high-quality chart editing data. Besides, we introduce two novel fine-grained, rule-based evaluation metrics: the layout metric, which evaluates the position, size and color of graphical components; and the text metric, which jointly assesses textual content and font styling.
Building on top of ChartEditVista, we present ChartEditor, a model trained using a reinforcement learning framework that incorporates a novel rendering reward to simultaneously enforce code executability and visual fidelity. Through extensive experiments and human evaluations, we demonstrate that ChartEditVista provides a robust evaluation, while ChartEditor consistently outperforms models with  similar-scale and larger-scale on chart editing tasks.
\end{abstract}

\section{Introduction}

\begin{figure*}[]
    \centering
     \includegraphics[width=1.0\linewidth]{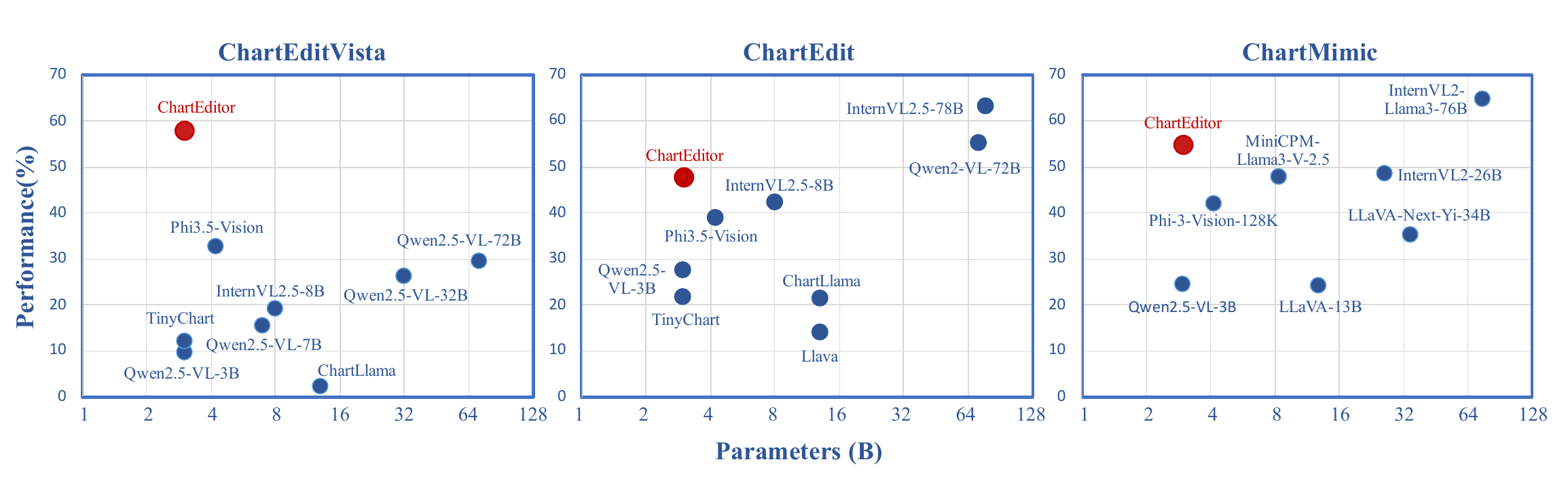}
    \caption{Comparisons on ChartEditor and other SOTA models, red dots represents ChartEditor and bule represent others.}
    \label{fig:intro}
\end{figure*}

Chart editing aims to utilize multi-modal large language models (MLLMs) to generate chart code to make high-quality visualizations reflect the customize requirements. This task is particularly practical as it significantly reduces the manual effort required for chart design and implementation. 

However, existing benchmarks ~\cite{yan2024chartreformer,chartedit,chartmimic} either assume complete chart code access or are limited in instruction diversity, editable elements, and chart types, with a lack of large-scale training data.

To address these limitations, we introduce ChartEditVista, a comprehensive benchmark for chart editing. ChartEditVista consists of 7,964 carefully curated samples with original chart images, natural language editing instructions, and corresponding modified code. It covers 31 chart types, 6 editing tasks and nearly all editable components. The benchmark is created through an automated pipeline combining scalable data generation and rigorous quality control, with multi-dimensional automated evaluation and expert manual refinement to ensure robust quality.

Our analysis also reveals limitations in existing metrics for chart editing tasks. MLLM-based metrics~\cite{chartedit,chartmimic} often exhibit hallucinations, while rule-based metrics~\cite{chartmimic} are inadequate for fine-grained modifications. To bridge this gap, we propose two novel fine-grained rule-based metrics: layout metric and text metric, which assess graphical elements and textual content respectively. Comprehensive human evaluations show strong alignment with human judgment.

Furthermore, our evaluation on ChartEditVista reveals that contemporary open-sourced MLLMs~\cite{qwenvl2series,internvl2series,llavaNext}, including chart specific models ~\cite{tinychart,chartllama,chartmoe,chartvlm_and_chartx}, show limited performance on chart editing tasks. 
To overcome these challenges, we present ChartEditor, a model specifically optimized for chart editing tasks. Built upon the Qwen2.5-VL-3B~\cite{qwen2_5series}, our model incorporates several key innovations. First, we introduce a novel rendering reward mechanism that directly assesses chart output fidelity, which serves as the optimization objective for Group Relative Policy Optimization (GRPO)~\cite{grpo,grpo_also}. Second, we implement a two-phase training framework combining supervised fine-tuning and curriculum reinforcement learning to stable reward signals. Third, we validate the generalizability of our rendering reward by successfully applying it to the Chart-to-Code task, where it yields consistent performance improvements.

Extensive experiments demonstrate that our ChartEditor-3B achieves state-of-the-art (SoTA) performance across MLLMs with similar scale on ChartEditVista while maintaining strong generalization on out-of-domain benchmarks including ChartEdit~\cite{chartedit} and ChartMimic~\cite{chartmimic}. Notably, it consistently outperforms a range of chart-specialized models and even surpasses several larger models with 26B and 34B parameters, highlighting the effectiveness of our model and training framework.

Our contributions are summarized as follows:

\begin{enumerate}
\item We introduce \textit{ChartEditVista}, a chart editing benchmark containing 7,964 manually verified samples spanning 31 chart types, covering diverse editable components and editing tasks. Our robust automated chart editing data generation pipeline, complemented by manual verification, ensures data quality.

\item We propose the \textit{layout} and \textit{text metric}, which outperform existing evaluation methods in reliability and fine-grained change detection. Human evaluations confirm their robustness and strong correlation with human assessments.

\item We present \textit{ChartEditor}, a model trained using GRPO and a novel \textit{rendering reward}. It achieves state-of-the-art performance on multiple benchmarks, including ChartEditVista, ChartEdit, and ChartMimic.

\end{enumerate}

\section{Related Works}
\textbf{Chart Related Works.} Charts, as a specific type of image used for data visualization, have garnered increasing research attention~\cite{chart_survey}. Existing works explore chart understanding from the perspectives of code~\cite{chartmimic,chartedit,chartvlm_and_chartx,AcademiaChart} and content~\cite{chart_survey,chartqa,chartvlm_and_chartx,chen2025chart,xcomposer,Xcomposer2,Xcomposer4k,IDEFICS2,mplugowl,mplug_oct,mplugowl2}. In the specific field, Chart Editing, significant progress has been made. For instance, some studies~\cite{yan2024chartreformer} have proposed workflow-based approaches for Chart Editing, while others~\cite{chartmimic,chartedit} have introduced multi-modal guidance to facilitate end-to-end chart modification. Despite these advancements, existing benchmarks still face limitations in terms of the diversity of instruction types, the variety of editable elements, and the range of chart types supported. Furthermore, Chart-specific Model~\cite{tinychart,chartllama,chartmoe,chartcoder} lack the ability of high-quality chart editing.

\noindent\textbf{Multimodal Large Language Models.} Recent research in Multimodal Large Language Models (MLLMs) has made substantial strides~\cite{Improved_baselines,visual_instruction_tuning,sphinx}. For example, closed-source MLLMs such as GPT-series~\cite{gpt4o,gpt4}, Claude-series~\cite{Claude3opus}, and Gemini-series~\cite{geminiprovision,gemini2_5} are capable of handling complex real-world vision and language tasks. Open-source MLLMs like LLaVA~\cite{llavaNext,llava}, Qwen-VL~\cite{qwen2_5series,qwenvl2series}, and InternVL-VL~\cite{internvl2series} also demonstrate strong performance compared to previous methods across several standard multimodal understanding benchmarks. However, these models still underperform when applied to tasks with high practical demands, such as Chart Editing. 

\section{ChartEditVista}

\subsection{Task Definition}

Chart editing involves generating a modified chart based on a given chart image and accompanying textual editing instructions. This task is commonly approached by predicting the underlying chart code that, once rendered, produces the desired edited chart. However, existing benchmarks often assume access to the original chart code, which is an unrealistic assumption in many real-world applications. To better align with practice, we formulate the task as follows:

\begin{equation}
C' = f(I, T),
\label{eq:task}
\end{equation}
where $I$ denotes the input chart image, $T$ denotes the editing instructions, $f$ is the model, and $C'$ is the code used to generate modified chart.

\begin{figure}
    \centering
    \includegraphics[width=1.0\linewidth]{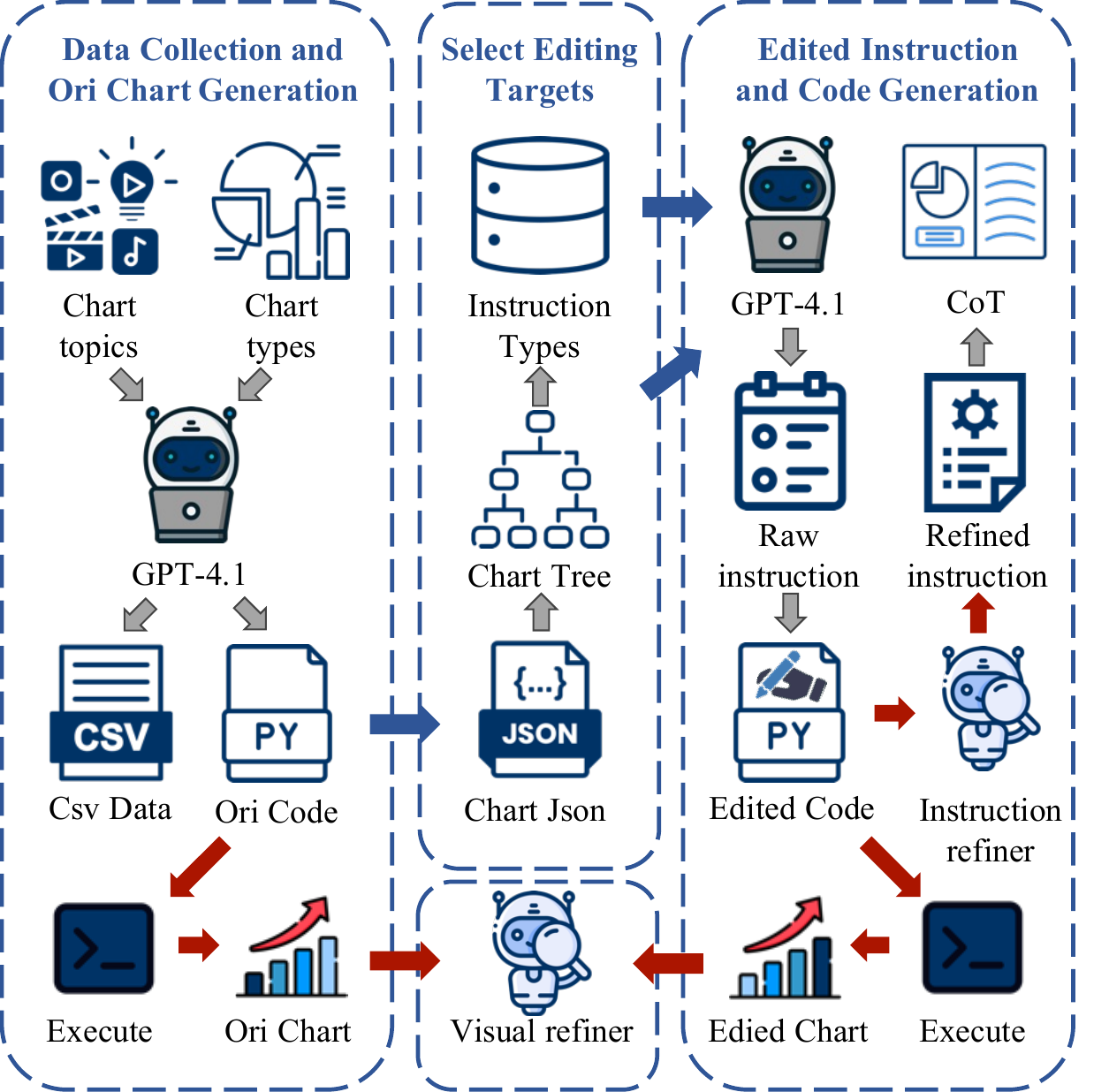}
    \caption{Overview of our fully data construction pipeline, it can generate chart editing data and refine them in various aspects. The gray arrows show the data flow inside each stage, the dark blue arrows show the data flow between stages and the red arrows show the data quality control steps.}
    \label{fig:data_pipeline}
\end{figure}

\subsection{Data Construction Pipeline}
\label{sec:data_pipeline}
To construct a diverse and semantically rich set on chart editing task, we adopt a multi-stage generation and filtering pipeline, as shown on Figure~\ref{fig:data_pipeline}. 

\noindent\textbf{Data Collection and Original Chart Generation.}
We begin by creating a chart topic pool and a chart type pool, aiming to ensure content diversity.
Chart topics are generated using GPT-4.1, yielding hundreds of unique topics. We manually predefine 31 chart types to ensure a broad range of visual formats.

Given a sampled chart topic and type, we employ GPT-4.1, selected for its strong code generation and visual reasoning capabilities~\cite{gpt4_1}, to generate synthetic data in CSV format and the corresponding Python plotting code, following the approach of ChartLlama~\cite{chartllama}.
To further enhance layout variability, we manually curated an attribute bank encompassing visual features such as the presence of multiple subplots, color schemes, axis configurations, and other stylistic properties. For each chart, we randomly sample a subset of attributes and incorporate them into the generation prompt to further enhance diversity.

All generated code is executed for initial quality control, with non-runnable scripts discarded. To further ensure visual clarity, each rendered chart is assessed by GPT-4.1 for interpretability and rendering quality (e.g., avoiding overlaps or truncation). Charts that fail this check are excluded, ensuring a pool of high-quality, visually coherent references.

\noindent\textbf{Selecting Editing Targets.}
To generate diverse and fine-grained chart editing instructions, it is crucial to first identify almost all potential editable elements within a chart. To achieve this, we utilize GPT-4.1 to convert the generated Python plotting code into a structured JSON format, referred to as the Chart JSON, which explicitly encodes all visual and textual elements present in the chart, along with their associated attribute values.

We then transform the Chart JSON into a hierarchical Chart Tree, where the root node represents the entire chart, internal nodes correspond to major chart components , and leaf nodes typically denote atomic elements. This tree structure enables systematic traversal for fine-grained editing and supports diverse instructions on both visuals and data.

For modify and delete operations, we randomly sample nodes from the Chart Tree as editing targets, ensuring broad coverage of editable entities. For add operations, we provide the entire Chart JSON as context to the model, enabling it to reason about where and how to insert new elements in a coherent manner.

\noindent\textbf{Editing Instructions and Code Generation.}
Given the Chart JSON and its corresponding Chart Tree, we instruct GPT-4.1 to perform two key tasks: (i) generate an initial editing instruction, and (ii) produce the corresponding edited chart code. If the sampled node is intrinsically non-editable, GPT-4.1 is instructed to skip that node and refrain from generating an instruction.

We apply a rigorous quality control protocol for the generated modify instructions and edited chart codes. As with reference charts, any edited code that fails to render a valid chart is discarded. Moreover, we prompt GPT-4.1 to refine the instruction based on the original and edited code, ensuring all the final instructions accurately and naturally describes the performed change. Finally, we submit the reference image, the refined instruction, and the edited image to GPT-4.1 for a holistic quality assessment. This check ensures that the editing outcome is not only syntactically correct but also interpretable and faithful to the modification.

To enhance the alignment between visual semantics and chart code, we generate a visual-code Chain-of-Thought (CoT) from the final code and its corresponding  refined instruction, inspired by ChartCoder~\cite{chartcoder}. Each CoT consists of three components:
(1) identification of the specific chart component(s) targeted by the instruction;
(2) specification of the attribute-level modifications required in the code; and
(3) description of the expected visual change resulting from the edited code.

\begin{table}[t]
\centering
\small
\setlength{\tabcolsep}{2pt}
\begin{tabular}{l c c c c c }
\toprule
\multirow{2}{*}{\textbf{Name}}  & \textbf{Chart} & \textbf{Data} & \textbf{Edit} & \textbf{Editable}  \\
 & \textbf{Types} & \textbf{number} & \textbf{type} & \textbf{Objects} \\
\midrule
ChartCraft~\cite{ChartCraft}      & 5 & 5.5k  & 4 & Limited \\
ChartX~\cite{chartvlm_and_chartx}                & 18 & 6k & 0 & None \\
ChartMimic~\cite{chartmimic}        & 22 & 2.4k & 1 & Limited  \\
ChartEdit~\cite{chartedit} & 19 & 1.4k & 6 & Limited  \\
\midrule
ChartEditVista (ours)   & 31 & 7.9k & 6 & Unlimited  \\
\bottomrule
\end{tabular}
\caption{Comparison of existing chart-related benchmarks. 
}
\label{tab:comparison_benchmarks}

\end{table}

\subsection{Benchmark Statistics and Analysis}
\label{sec:charteditbench}

\textbf{Data Statistics.}
Building on the aforementioned automatic pipeline, we construct ChartEditVista, a comprehensive benchmark comprising \textit{601} unique base charts and a total of \textit{7{,}964} triplets in the form of \textit{(chart, instruction, modified code)}. 

\noindent\textbf{Diversity Analysis.} As shown in Table \ref{tab:comparison_benchmarks} and Appendix A, ChartEditVista covers \textit{31} chart types. It spans \textit{six} major instruction categories—\textit{add visual components}, \textit{modify visual components}, \textit{delete visual components}, \textit{add data},\textit{modify data} and \textit{delete data}. It maintains a balanced distribution of single- (53.4\%) and multi-subplot (46.6\%) charts, and exhibits substantial variation in instruction complexity, with single-instruction cases accounting for 35.8\% of samples and multi-instruction cases ranging up to six steps. Furthermore, the editing granularity ranges from fine-grained modifications of individual component attributes to global edits affecting the entire chart, supported by the hierarchical Chart Tree representation.

\noindent\textbf{Quality Analysis.} To ensure data quality, we engaged 10 trained annotators to manually validate all 7,964 triplets. The overall instruction clarity exceeded 97\%, and the success rate of visually implementing modification instructions in the charts was over 94\%. For the visual quality assessment of the original charts, annotators rated image clarity, visual occlusion, and subplot layout on a three-point scale. After normalization, the average scores for these criteria exceeded 96\%, 91\%, and 98\%, respectively, more details are in Appendix A.

\begin{table}[t]
\centering
\scriptsize
\setlength{\tabcolsep}{1pt}
\renewcommand{\arraystretch}{1.05}
\begin{tabular}{l|cccc}
\toprule
\multirow{2}{*}{Benchmark} & \multicolumn{2}{c}{\textbf{Text}} & \multicolumn{2}{c}{\textbf{Layout}} \\
\cmidrule(lr){2-3}\cmidrule(lr){4-5}
 & $r$ & $\rho$ & $r$ & $\rho$ \\
\midrule
ChartEditVista & 0.779\,(1.4e$^{-21}$) & 0.860\,(2.5e$^{-30}$) & 0.712\,(2.1e$^{-16}$) & 0.718\,(4.0e$^{-17}$) \\
ChartMimic    & 0.775\,(2.6e$^{-21}$)  & 0.790\,(1.6e$^{-22}$)  & 0.849\,(5.4e$^{-29}$) & 0.847\,(1.3e$^{-28}$) \\
ChartEdit   & 0.778\,(1.8e$^{-21}$)  & 0.739\,(1.4e$^{-18}$)  & 0.783\,(5.4e$^{-22}$) & 0.800\,(1.8e$^{-23}$) \\
\bottomrule
\end{tabular}
\vspace{-0.5em}
\caption{The correlations observed on three chart editing benchmarks indicate that our metrics exhibit strong and statistically significant agreement with human evaluation.}
\label{tab:metric_human_eval}
\vspace{-2.0em}
\end{table}

\subsection{Fine-grained Evaluation Metrics}

\begin{figure}[t]
    \centering
    \includegraphics[width=1.0\linewidth]{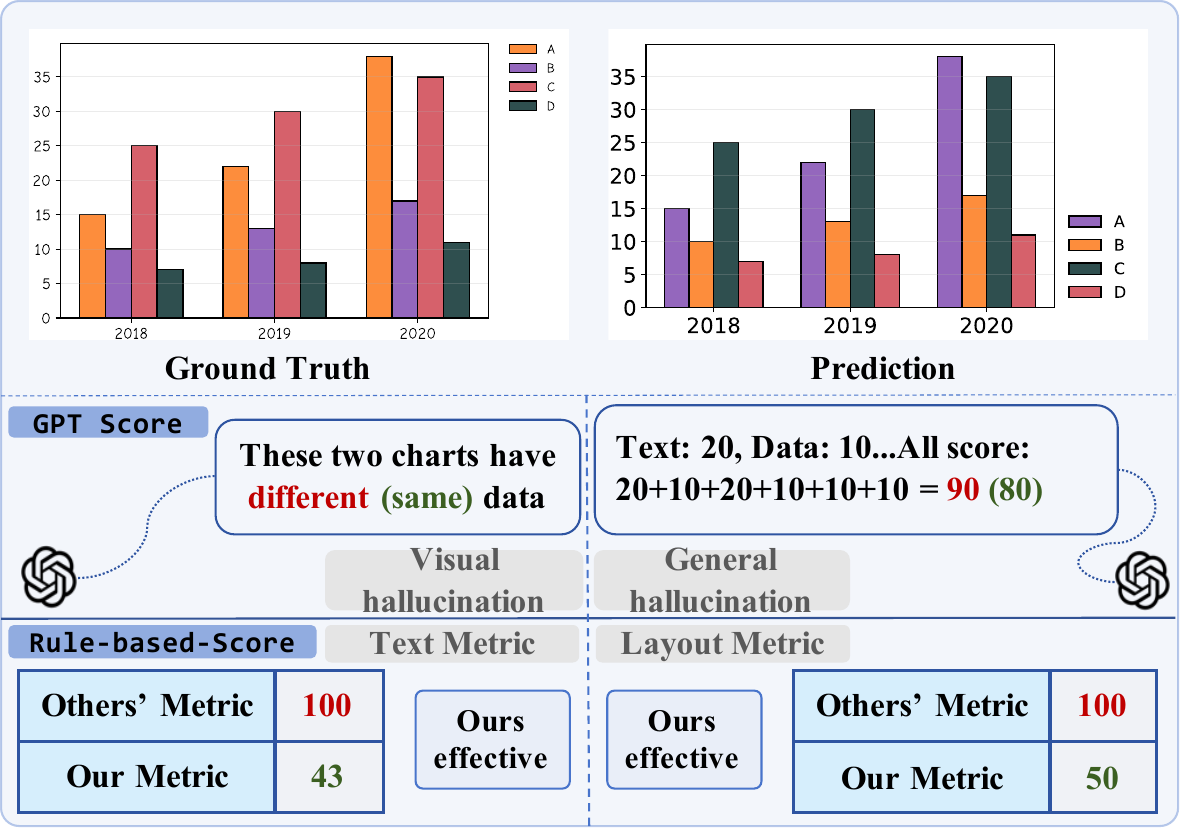}

    \caption{Limitations of  existing metrics, while MLLM-Based metric suffers from serious hallucination, rule-based metric can not capture the fine-grained visual changes.}
    \label{fig:metric}

\end{figure}

As shown in Figure~\ref{fig:metric} and Appendix A, MLLM-based metrics often hallucinate, while existing rule-based metrics are limited to coarse, subplot-level evaluation and cannot capture fine-grained edits. To address these shortcomings, we propose Rendering-Aware Rule-based Metrics (RARM), which assess visual similarity based on color, position, and style attributes of chart components. RARM comprises two complementary metrics: a layout metric for general graphical objects and a text metric for textual elements.

For graphical objects, similarity is defined as:
\begin{equation}
S_{L}(p, g) = S_{\mathrm{color}}(p, g) \times S_{\mathrm{pos}}(p, g) \times S_{\mathrm{shape}}(p,g),
\end{equation}

where color similarity is assessed via normalized Euclidean distance in RGB space and positional/shape similarity follow type-specific definitions (e.g., center distance and aspect ratio for patch-like objects).

For text elements, $S_{\mathrm{base}}(p, g)$ is defined as 1 if the textual content matches exactly, and 0 otherwise. The final text similarity is computed as
\begin{equation}
S_{T}(p, g) = S_{\mathrm{base}}(p, g) \cdot (1 - \lambda M_f - \alpha M_s),
\end{equation}
where $M_f$ and $M_s$ indicate font family and size mismatches, respectively, and $\lambda, \alpha$ are penalty coefficients (empirically set to 0.3).

Finally, we compute overall layout and text metrics using optimal assignment via the Hungarian algorithm, enabling robust and fine-grained evaluation of chart edits.

To assess reliability, we evaluated our metrics against human judgments on three chart editing benchmarks. As shown in Table~\ref{tab:metric_human_eval}, both text and layout metrics achieve Pearson and Spearman correlations exceeding 0.7 ($p \ll 0.01$) across all datasets, confirming strong and statistically significant alignment with human evaluation.

\section{Method}

\begin{figure*}[t]
    \centering
    \includegraphics[width=1.0\linewidth]{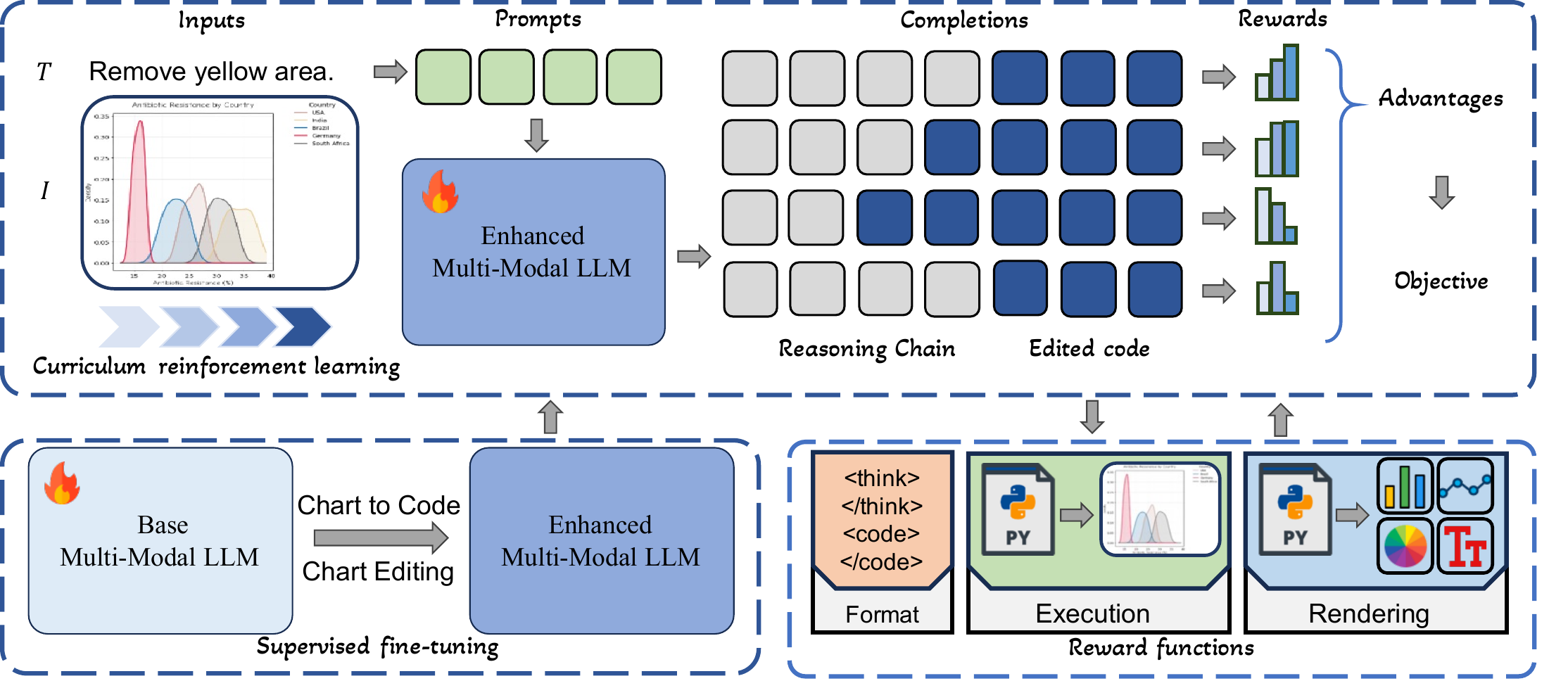}

    \caption{Overview of ChartEditor’s training procedure. ChartEditor is optimized within a GRPO-based reinforcement learning framework, guided by format, execution, and rendering rewards. To further stabilize learning signals, ChartEditor employs a multi-stage training schema that incorporates supervised fine-tuning for cold-start and curriculum RL strategies.}
    \label{fig:training}

\end{figure*}

We begin by introducing the GRPO algorithm. Subsequently, we detail our cold-start strategy, which systematically enhances the base model’s chart editing capabilities. Finally, we present our design of multiple reward functions and the curriculum reinforcement learning procedure. More details are available in Appendix B.

\subsection{Preliminary}
\label{preliminary}
Group Relative Policy Optimization (GRPO) is a reinforcement learning (RL) algorithm designed to optimize language models efficiently without requiring an explicit value model. GRPO modifies the standard Proximal Policy Optimization (PPO) objective by computing token-level advantages based on group-normalized rewards.

During reinforcement learning, 

the policy model $\pi_\theta(\cdot \mid \cdot)$ generates a batch of $N$ rollouts $\{o_i\}_{i=1}^G$, and optimize towards the rollout that achieve higher scores.

\subsection{Multi-Stage Cold-Start}
\label{sft}

Directly applying MLLMs to RL framework is suboptimal due to two key limitations in chart editing tasks. First, MLLMs struggle with aligning visual chart content with their corresponding plotting code, making it difficult to accurately map visual elements to code representations. Second, they exhibit weak associations between natural language editing instructions and the necessary code modifications, limiting their ability to precisely translate instructions into code changes. These misalignments lead to sparse and noisy reward signals during RL training, reducing both efficiency and downstream performance.
To mitigate these issues, we propose a systematic two-stage supervised fine-tuning (SFT) pipeline using the Qwen-2.5-VL 3B model to improve cross-modal alignment.

\noindent\textbf{Stage 1: Chart-to-Code SFT.}
The objective of this stage is to improve the correspondence between chart visual features and plotting code. The model is fine-tuned on a large-scale Chart-to-Code corpus, with the introduction of visual-to-code chain-of-thought (CoT) explanations to enable explicit reasoning over the mapping between image elements and code statements.

\noindent\textbf{Stage 2: Chart Editing SFT.}
The goal of this stage is to strengthen the alignment between editing instructions and code modifications. The model undergoes further fine-tuning with chart editing data, incorporating editing-specific CoT demonstrations that decompose instructions, specify required code edits, and describe the expected visual outcomes, thereby reinforcing the model’s capacity for instruction interpretation and code adaptation.

This two-stage SFT process facilitates integration chart visual representations, natural language editing instructions, and modifiable code components. By establishing these strengthened cross-modal alignments, the approach provides a robust foundation for subsequent reinforcement learning, ultimately leading to enhanced chart editing performance.

\subsection{Reward Functions}
\label{rewards}

\textbf{Format Reward.}
We prompt the MLLM to place its reasoning trace inside \texttt{<think>}~\dots~\texttt{</think>} tags and the final code inside \texttt{<code>}~\dots~\texttt{</code>} tags. During training, we parse every rollout and attempt to extract both segments; if the reasoning trace and the code snippet are successfully recovered, the \emph{Format Reward} is set to 1, otherwise 0.

\noindent\textbf{Execution Reward.}
For every rollout produced by the MLLM, we run the extracted final code inside an isolated sandbox. Each sandbox is independent, ensuring that different executions cannot interfere with one another. If the code terminates successfully within a predefined time limit, the \emph{Execution Reward} $R_\mathrm{E}$ is assigned a value of 1; if it times out or raises an error, the code is deemed non‑executable and the reward is set to 0.

\noindent\textbf{Rendering Reward.}
To faithfully capture the correspondence between code and visual output, we introduce a rule-based \emph{Rendering Reward}. For each object type, we compute a pairwise similarity matrix $S \in \mathbb{R}^{M \times N}$ between $M$ predicted and $N$ ground truth objects:
\begin{equation}
S_{j,k} = \operatorname{sim}(obj_j^{\text{pred}}, obj_k^{\text{gt}})
\end{equation}
where $\operatorname{sim}(\cdot, \cdot)$ is a type-specific similarity metric (e.g. IoU for patch-like objects, $L_2$ distance for point objects). We then solve an optimal assignment problem using the Hungarian algorithm and normalize by the maximum object count:
\begin{equation}
S_\mathrm{type} = \frac{1}{\max(M, N)} \sum_{(j, k) \in \mathcal{M}^*} S_{j, k}
\end{equation}
where $\mathcal{M}^*$ denotes the optimal matching. The final rendering reward aggregates the normalized scores across all object types, weighted as appropriate, and multiplied by the executability indicator $R_\mathrm{E}$:
\begin{equation}
\mathcal{R}_\mathrm{render} = R_\mathrm{E} \sum_{t \in \mathcal{T}} w_t\, S_{\text{type}=t}
\end{equation}
where $R_\mathrm{E} \in {0,1}$ denotes code executability, more details are in Appendix C.

\begin{table*}[t]

\centering
\setlength{\tabcolsep}{3pt}
\begin{small}
\small
\resizebox{\linewidth}{!}{
\begin{tabular}{l r | cccccc | ccc | cccc}
\toprule
\multirow{2}{*}{Model} & \multirow{2}{*}{\textbf{\#~}Params.} &
\multicolumn{6}{c|}{ChartEditVista} & \multicolumn{3}{c|}{ChartEdit \textit{w/o} Code} & \multicolumn{4}{c}{ChartMimic Customized} \\
\cmidrule(lr){3-8} \cmidrule(lr){9-11} \cmidrule(lr){12-15}
& & Exec. & $\mathrm{T_R}$& $\mathrm{L_R}$ & Col. & Type & Avg. &
Exec. & Code & Chart &
Exec. & Low & High & Overall \\
\midrule
\multicolumn{15}{c}{\textit{Proprietary Models}}\\
\midrule
GPT\!-4o~\cite{gpt4o} & -- & 69.3 & 31.5 & 48.9 & 52.3 & 41.2 & 43.5 &
91.5 & 60.0 & 79.9 &
\textbf{96.5} & \textbf{82.1} & \textbf{84.3} & \textbf{83.2}\\

Claude-3.7-Sonnet~\cite{claude37} &-- & \textbf{94.0} & 46.2 & 55.9 & 64.1 &  75.8 & 60.5 &
\textbf{96.3} & 73.4 & 87.5 &
80.6 & 67.4 & 78.1 & 72.3\\

Gemini-2.5-Pro~\cite{gemini2_5} & -- &90.5& \textbf{55.1} & \textbf{63.4} & \textbf{69.0}& \textbf{76.3} & \textbf{66.0} &
95.3 & \textbf{82.9} & \textbf{89.2} &
91.7 & 81.3 & 83.5 & 82.4 \\

\midrule
\multicolumn{15}{c}{\textit{Open-Source Models}}\\
\midrule
Qwen2.5-VL-72B~\cite{qwen2_5series} & 72B & 80.6 & 11.3 & 39.4 & 38.5 & 30.0 & 29.8 &
89.8 & 57.7 & 71.0 &
85.3 & 67.6 & 69.1 & 68.4\\

Qwen2.5-VL-32B~\cite{qwen2_5series} & 32B & 74.7 & 9.1 & 35.6 & 33.9 & 26.6 & 26.3 &
88.4 & 61.7 & 72.5&
85.6 & 66.4 & 69.5 & 68.0\\
Qwen3-VL-32B~\cite{Qwen3-VL}  & 32B & 60.6 &28.5 & 47.5 & 41.4 & 45.7 & 40.8 & 
 84.9& 69.0 & 73.4 & 
88.2&72.4 & 61.3 & 66.9 \\

LLaVA-1.5-13B~\cite{llava} & 13B & -- & -- & -- & -- & -- & -- &
48.8 & 11.4 & 16.7 &
49.0 & 23.6 & 24.7 & 24.2\\
ChartLlama~\cite{chartllama} & 13B &19.3 & 2.1 & 7.3 & 4.7 & 3.7 & 4.5 &
52.3 & 15.8 & 27.4 &
-- & -- & -- & --\\
Qwen3-VL-8B~\cite{Qwen3-VL} & 8B & 61.9 &26.4 & 46.3 & 37.7 & 45.4 & 39.0 & 
 84.9& 59.6 & 69.3 & 
78.1& 61.4& 60.8 & 61.1 \\
Qwen2.5-VL-7B~\cite{qwen2_5series} & 7B & 57.1 & 6.1 & 14.4 & 18.9  & 22.3& 15.4 &
65.1 & 43.4 & 44.0  &
68.4 & 46.0& 70.0 & 58.0 \\

\midrule

Qwen3-VL-4B~\cite{Qwen3-VL}  & 4B & 60.0 &19.2 & 39.0 & 30.0 & 34.0 & 30.6 & 
\textbf{79.5} & \textbf{55.6} &  \textbf{63.2}& 
\textbf{66.0}& 47.6& 49.1& 48.4 \\
Qwen2.5-VL-3B~\cite{qwen2_5series} & 3B & 45.8 & 6.2 & 10.3 & 9.5 & 11.8 & 9.5 &
44.1 & 31.1 & 24.3 &
48.8 & 26.2 & 22.9 & 24.6\\
TinyChart~\cite{tinychart} & 3B &30.1 & 6.2 & 13.8 & 13.8 & 13.0 & 12.2 &
36.3 & 18.3 & 25.1 &
-- & -- & -- & --\\
Qwen3-VL-2B~\cite{Qwen3-VL}  & 2B & 47.6 &16.6 & 29.7 & 25.3 & 28.9 & 25.1 & 
 67.5& 36.7 & 44.4 & 
56.8& 37.8&37.3 & 37.6 \\

\midrule
\cellcolor[HTML]{efefef}{ChartEditor (\textbf{Ours})} & \cellcolor[HTML]{efefef}{3B} &  \cellcolor[HTML]{efefef}{\textbf{76.8}} &  \cellcolor[HTML]{efefef}{\textbf{52.9}} &  \cellcolor[HTML]{efefef}{\textbf{48.7}} &  \cellcolor[HTML]{efefef}{\textbf{64.1}} &  \cellcolor[HTML]{efefef}{\textbf{67.5}}  &  \cellcolor[HTML]{efefef}{\textbf{58.1}} &
\cellcolor[HTML]{efefef}{66.5} & \cellcolor[HTML]{efefef}{40.2} & \cellcolor[HTML]{efefef}{55.3} &
\cellcolor[HTML]{efefef}{61.6} & \cellcolor[HTML]{efefef}{\textbf{52.1}} & \cellcolor[HTML]{efefef}{\textbf{57.9}} & \cellcolor[HTML]{efefef}{\textbf{55.0}}\\
\bottomrule

\end{tabular}
}
\caption{
Performance on  ChartEditVista (\textbf{in-domain}) and ChartEdit w/o code and ChartMimic Customized (\textbf{out-of-domain}). }
\label{tab:combined_results}
\end{small}
\end{table*}

\subsection{Training}
\label{training}
\textbf{Curriculum Reinforcement Learning.}
In the initial training phase, the model rarely produces executable code, resulting in highly sparse reward signals. 
To facilitate the training,  
we implement a progressive curriculum reinforcement learning strategy: initial training on single-subplot, single-instruction cases; followed by gradual introduction of multi-subplot and multi-instruction examples. This approach maintains stable reward density throughout training.

\noindent\textbf{Reward Computation.}
For each rollout, we compute two sub-rewards: a format reward, and a rendering reward depends on execution reward, reflecting code structure, executability, and visual fidelity, respectively. The final reward is defined as the summation of these sub-rewards:
\begin{equation}
r_i = r^{\text{format}}_i +r^{\text{render}}_i
\end{equation}

\noindent\textbf{Policy Update.}
The rewards $\{ r_i \}_{i=1}^G$ are normalized into advantages via Z-score normalization,
Finally, the model is optimized by minimizing the objective:
\begin{equation}
L = -\frac{1}{G} \sum_{i=1}^{G} \frac{\pi(o_i \mid I,T)}{\pi_{\mathrm{old}}(o_i \mid I,T)} \cdot \hat{A}_{i,t}
\end{equation}
where $\pi_{\mathrm{old}}(\cdot \mid \cdot)$ denotes the old policy and $\hat{A}_{i,t}$ is the advantage associated with prediction $o_i$.

\section{Experiments}
\subsection{Experimental Settings}

\textbf{Datasets.}
\label{subsec:dataset}
After filtering out examples containing stochastic code statements such as \texttt{random.choice},we select 140k high-quality samples from the ChartCoder-160k~\cite{chartcoder} corpus for stage-1 supervised fine-tuning (SFT). We then perform stage-2 SFT with 20k instances constructed by our chart editing data generation pipeline, and further train the model with Generalized Reward Policy Optimization (GRPO)~\cite{grpo_also} on 6k additional ChartEditVista samples.We report results on two benchmark suites:
(i) the \textit{in-domain} ChartEditVista, and  
(ii) the \textit{out-of-domain}, we select ChartEdit~\cite{chartedit} and ChartMimic-Customized~\cite{chartmimic} test sets because they are designed with the absence of code input in mind.


\noindent\textbf{Implementation Details.}
\label{subsec:impl}We adopt Qwen-2.5-VL-3B~\cite{qwen2_5series} as the default backbone. All SFT is conducted with a batch size of 16, an initial learning rate of 1e-5, and a weight decay of 0.01. For GRPO, we use a batch size of 2 and generate 8 rollout samples per training step. The execution-timeout threshold in the reward function is set to 30\,s. Optimizer hyper-parameters remain identical to SFT (learning rate 1e-5, weight decay 0.01). We fully fine-tune all parameters of the MLLM throughout training.

\noindent\textbf{Evaluation Metrics.} 
We evaluate ChartEditVista using our proposed RARM layout $\mathrm{L_R}$ and text $\mathrm{T_R}$ metrics to provide fine-grained assessment. For overall chart type and color evaluation, we follow the same evaluation protocols as ChartMimic~\cite{chartmimic}. On the ChartEdit and ChartMimic benchmarks, we report results using each benchmark's official metrics to ensure fair comparison. 

\subsection{Results}

\textbf{Main Results.} As shown in Table~\ref{tab:combined_results}, our 3B \textbf{ChartEditor} sets a new state-of-the-art for open-source models on the in-domain ChartEditVista benchmark with an average score of 58.1. It significantly surpasses both prior specialized models like TinyChart (12.2) and much larger general-purpose VLMs like Qwen2.5-VL-72B (29.8). Notably, ChartEditor also outperforms the proprietary GPT-4o, both in average score (58.1 vs. 43.5) and execution accuracy (76.8 vs. 69.3), demonstrating that a compact, domain-specialized model can outperform significantly larger systems.
Despite being trained on synthetic data, ChartEditor exhibits strong out-of-domain generalization. It achieves competitive scores of 55.3  on ChartEdit w/o Code and 55.0 on ChartMimic Customized, outperforming many larger models like LLaVA-1.5-13B (16.7). On the latter benchmark, it also achieves the best low-level and high-level scores among similarly-sized open-source models. More detailed results and error analysis are available in Appendix D.

\setlength{\tabcolsep}{8pt}{
\begin{table}[t]
\centering
\small

\setlength{\tabcolsep}{6.3pt}
\begin{tabular}{cc|cccccc}
\toprule
\multicolumn{2}{c|}{SFT}  &
\multicolumn{6}{c}{Metrics} \\
\cmidrule(lr){1-2}\cmidrule(lr){3-8}
C2C & Edit  &
Exec. & $\mathrm{T_R}$ &$\mathrm{L_R}$ & Type & Col. & Avg. \\
\midrule
-- & --            & 47.2 & 14.2 & 11.2 & 25.7 & 22.5 & 18.5 \\
-- & $\checkmark$  & \textbf{77.8} & 45.6 & 43.9 & 63.7 & 59.8 & 53.2 \\
$\checkmark$ & --  & 73.4 & 45.9 & 46.1 & 64.3 & 60.7 & 54.3 \\
$\checkmark$    & $\checkmark$ & 76.8 & \textbf{52.9} & \textbf{48.7} & \textbf{67.5} & \textbf{64.1} & \textbf{58.1} \\  
\bottomrule

\end{tabular}
\caption{
Ablation on the training strategy without curriculum learning.
}
\label{tab:ablation_train_nocurr}
\end{table}}

\begin{table}[t]
\centering
\small
\setlength{\tabcolsep}{7.45pt}
\begin{tabular}{c|cccccc}
\toprule
Curric. &
Exec. & $\mathrm{T_R}$ & $\mathrm{L_R}$ & Type & Col. & Avg. \\
\midrule
--              & 73.6 & \textbf{53.0} & 46.8 & 65.1 & 61.3 & 56.6 \\  
$\checkmark$    & \textbf{76.8} & 52.9 & \textbf{48.7} & \textbf{67.5} & \textbf{64.1} & \textbf{58.1} \\ 
\bottomrule

\end{tabular}

\caption{
Ablation on curriculum learning for Chart-to-Code and Chart-Editing data with RL.
}

\label{tab:ablation_curric}
\end{table}

\begin{table}[t]
\centering
\small
\setlength{\tabcolsep}{5.8pt}
\renewcommand{\arraystretch}{1.15}
\begin{tabular}{ccc|cccccc}
\toprule
\multicolumn{3}{c|}{Rewards} &
\multicolumn{6}{c}{Metrics} \\
\cmidrule(lr){1-3}\cmidrule(lr){4-9}
\textbf{R} & \textbf{E} & \textbf{F} & Exec. &$\mathrm{T_R}$& $\mathrm{L_R}$ & Type & Col. & Avg. \\
\midrule
--           & --           & $\checkmark$ & 66.3 & 47.6 & 43.3 & 58.0 & 54.3 & 50.8 \\
--           & $\checkmark$ & $\checkmark$ & 73.8 & 46.5 & 47.0 & 64.3 & 53.7 & 54.6 \\
$\checkmark$ & $\checkmark$ & --           & 75.3 & \textbf{53.4} & 47.8 & 66.0 & 61.5 & 57.2 \\
$\checkmark$ & $\checkmark$ & $\checkmark$ & \textbf{76.8} & 52.9 & \textbf{48.7} & \textbf{67.5} & \textbf{64.1} & \textbf{58.1} \\
\bottomrule
\end{tabular}

\caption{
Ablation on reward components. R: rendering, E: execution, F: format.
}
\label{tab:reward_ablation_chk}
\end{table}

\begin{table}[t]
\centering

\setlength{\tabcolsep}{2pt} 
\renewcommand{\arraystretch}{1.05} 
\begin{small}
\begin{tabular}{l r c c c c}
\toprule
Model & Exec. & Low & High & Overall \\ 
\midrule

GPT-4o~\cite{gpt4o}        & \textbf{93.2} & \textbf{79.0} & \textbf{83.5} & \textbf{81.2} \\
GeminiProVision~\cite{geminiprovision}       & 68.2 & 53.8 & 53.3 & 53.6 \\
Claude-3-opus~\cite{Claude3opus}        & 83.3 & 60.5 & 60.1 & 60.3 \\
\midrule

InternVL2-8B~\cite{internvl2series}             & 61.8 & 34.4 & 38.9 & 36.6 \\

Qwen2-VL-7B~\cite{qwenvl2series}              & 67.0 & 32.9 & 35.0 & 34.0 \\
LLaVA-Mistral-7B~\cite{llava}       & 59.7 & 20.7 & 21.3 & 21.0 \\
InternVL2-4B~\cite{internvl2series}            & 66.2 & 33.8 & 38.4 & 36.1 \\
Qwen2.5-VL-3B~\cite{qwen2_5series}           & 49.3 & 35.2 & 17.0  & 26.1   \\
Qwen2.5-VL-3B-SFT  & 49.3 & 35.2 & 17.0  & 26.1 \\
Qwen2.5-VL-3B-Render           & \textbf{69.6} & \textbf{52.7} & \textbf{56.4} & \textbf{54.5} \\
\bottomrule
\end{tabular}

\caption{Results on ChartMimic Direct Mimic task.
}

\label{tab:main_results_direct_compact}
\end{small}
\end{table}

\subsection{Ablation Study}

In this section, we provide a set of comprehensive ablations to validate the efficiency of ChartEditor, furthermore, we provide significance-test and task ablation in Appendix D.

\noindent\textbf{Impact of SFT Data Sources.}
As shown in Table~\ref{tab:ablation_train_nocurr}. We ablate the SFT data while keeping curriculum learning and RL fixed. Without SFT, the model performs poorly (47.2 Exec, 18.5 Avg). Training only on \textit{ChartEditing} data significantly boosts execution to 77.8 and average to 53.2. Using only \textit{Chart2Code} yields comparable execution (73.4) and slightly better fine-grained metrics (54.3 Avg), showing its complementary effect. Combining both achieves the best results (76.8 Exec, 58.1 Avg), confirming that structural and edit-specific data provide additive benefits.

\noindent\textbf{Effect of Curriculum Learning.}
With RL, Chart2Code, and ChartEditing data fixed, introducing curriculum learning increases execution accuracy from 73.6 to 76.8 and improves the overall score from 56.6 to 58.1, as shown in Table~\ref{tab:ablation_curric}. The main improvements are observed in the visual-related metrics, including Layout (increase of 1.9), Type (increase of 2.4), and Color (increase of 2.8), while the Text score remains nearly unchanged. These results indicate that progressive curriculum learning stabilizes training and leads to more accurate and visually consistent chart editing.

\noindent\textbf{Reward Design.}
As shown in Table~\ref{tab:reward_ablation_chk}, we analyze the effect of progressively adding reward components: format (\textbf{F}), execution (\textbf{E}), and rendering (\textbf{R}). Starting with only the format reward, the model achieves 50.8 average score and 66.3\% execution accuracy, indicating limited capability without execution guidance. Adding the execution reward (E + F) significantly improves performance to 54.6 average and 73.8 execution, reflecting the crucial role of executable correctness. Incorporating the rendering reward instead of format (R + E) further boosts results to 57.2 average and 75.3 execution, mainly enhancing visual-related metrics. Finally, combining all three rewards yields the best overall results, with 58.1 average and 76.8 execution accuracy. The results indicate that the execution reward ensures code correctness, the rendering reward enhances visual fidelity, and the format reward enforces output consistency and parsability. Combining these rewards yields complementary improvements in overall model performance. To further evaluate the applicability of rendering reward, we apply it to chart-to-code task, which denotes Qwen2.5-VL-3B-Render, as shown in Table\ref{tab:main_results_direct_compact}, our rendering reward can also achieve comparable performance on ChartMimic Direct Mimic, demonstrate the benefits of  rendering reward in chart-code related tasks.

\section{Conclusion}

We present ChartEditVista, a comprehensive chart editing benchmark with broad coverage and high-quality annotations, together with two rendering-aware evaluation metrics that robustly capture fine-grained visual and textual modifications. Leveraging these resources, our ChartEditor model, trained via GRPO with a novel rendering reward and curriculum learning, achieves state-of-the-art results on both in-domain and out-of-domain benchmarks. These contributions establish a new foundation for the rigorous evaluation and development of MLLMs for chart editing.

\section{Acknowledgments}
This work was supported by the Beijing Natural Science Foundation (No. L233008).

\bibliography{aaai2026}

\appendix

\end{document}